\documentclass[prd,twocolumn,superscriptaddress,floatfix,nofootinbib]{revtex4-2}
\pdfoutput=1 
\usepackage[T1]{fontenc}
\usepackage{amsmath}
\usepackage{amssymb}
\usepackage{amsfonts}
\usepackage{graphicx}
\usepackage{enumitem}
\usepackage{bm}
\usepackage{rotating}
\usepackage{hyperref}
\usepackage{pbox}
\usepackage[dvipsnames]{xcolor}
\usepackage{array}
\usepackage{physics}
\usepackage{dsfont}
\usepackage{mathtools}
\usepackage{leftidx}
\usepackage{lmodern}
\usepackage[normalem]{ulem}
\usepackage{braket}
\usepackage{tensor}
\usepackage{mathrsfs}
\usepackage{float}
\usepackage{dsfont}
\usepackage[margin=2cm]{geometry}
\usepackage[title]{appendix}
\usepackage{tcolorbox}

\usepackage{tikz}
\usetikzlibrary{arrows.meta,decorations.pathmorphing,math}

\newcommand{\tcb}[1]{\leavevmode{\color{Blue}{#1}}}

\newcommand{\sx}{\mathsf{x}}
\newcommand{\sy}{\mathsf{y}}
\newcommand{\sz}{\mathsf{z}}

\newcommand{\bx}{{\bm{x}}}
\newcommand{\bk}{{\bm{k}}}
\newcommand{\id}{\mathds{1}}
\newcommand{\ii}{\mathsf{i}}

\newcommand{\ts}[1]{_\textsc{#1}}

\newcommand{\kako}[1]{\left( #1 \right)}
\newcommand{\kagikako}[1]{\left[ #1 \right]}
\newcommand{\Bigkako}[1]{\Big( #1 \Big)}

\newcommand{\Bigkagikako}[1]{\Big[ #1 \Big]}

\newcommand{\R}{\mathbb{R}}
\newcommand{\C}{\mathbb{C}}
\newcommand{\M}{\mathcal{M}}
\newcommand{\A}{\mathcal{A}}
\newcommand{\W}{\mathcal{W}}
\newcommand{\CS}{C^\infty_0(\M)}
\newcommand{\Sol}{\mathsf{Sol}}

\newcommand{\Dab}{\Delta_\textsc{ab}}

\begin{document}

\title{Channel capacity of relativistic quantum communication with rapid interaction}

\author{Erickson Tjoa}
\email{e2tjoa@uwaterloo.ca}
\affiliation{Department of Physics and Astronomy, University of Waterloo, Waterloo, Ontario, N2L 3G1, Canada}
\affiliation{Institute for Quantum Computing, University of Waterloo, Waterloo, Ontario, N2L 3G1, Canada}

\author{Kensuke Gallock-Yoshimura}
\email{kgallock@uwaterloo.ca}
\affiliation{Department of Physics and Astronomy, University of Waterloo, Waterloo, Ontario, N2L 3G1, Canada}

\begin{abstract}

In this work we study nonperturbatively the transmission of classical and quantum information in globally hyperbolic spacetimes, where the communication channel is between two qubit detectors interacting with a quantized massless scalar field via delta-coupling interaction. This interaction approximates very rapid detector-field interaction, effectively occurring at a single instant in time for each detector. 
We show that when both detectors interact via delta-coupling, one can arrange and tune the detectors so that the channel capacity is (at least) as good as the quantum channel constructed nonperturbatively using \textit{gapless detectors} by Landulfo [\href{https://journals.aps.org/prd/abstract/10.1103/PhysRevD.93.104019}{\tcb{PRD \textbf{93}, 104019}}]. Furthermore, we prove that this channel capacity is in fact optimal, i.e., both nonperturbative methods give  essentially the same channel capacity, thus there is a sense in which the two methods can be regarded as equivalent as far as relativistic quantum communication is concerned. 

\end{abstract}

\maketitle
\flushbottom

\section{Introduction}

A central question in quantum information theory is how much (classical or quantum) information that can be transmitted between two or more parties via a quantum communication channel. Since quantum information theory is itself agnostic to the underlying spacetime geometry and states of motion of the relevant parties, relativistic effects must therefore enter through the types of quantum channel used to establish the communication. Early results that demonstrate the role of relativistic fields on quantum communication (based on Bogoliubov-transformation arguments) can be seen in, e.g., \cite{mancini2014preserving,Good2021communicationmirror,gianfelici2017quantum}. In particular, a fully covariant \textit{relativistic quantum communication} (RQC) protocol must therefore involve quantum fields in general curved spacetimes, as well as two or more observers each carrying a localized quantum system (such as a qubit or an atom) used as part of the communication protocol. The states of motion of these observers must also be described in a covariant and relativistically consistent manner. 

Indeed, such description can be provided by the Unruh-DeWitt (UDW) particle detector model, where the detectors are pointlike two-level systems interacting locally with a quantized scalar field \cite{Unruh1979evaporation,DeWitt1979}. The UDW model provides a simplified model of light-matter interaction where the exchange of angular momentum is not important (since we are throwing away vectorial nature of typical interactions between, say, an electromagnetic field with atomic dipole). The UDW model has been refined to the extent that it now admits fully covariant description that includes finite-size effect \cite{Tales2020GRQO,Bruno2021broken} and quantized centre of mass degrees of freedom \cite{Lopp2021deloc}. The UDW model also allows for local measurement theory \cite{polo2021detectorbased} and simplified generalizations that can capture effects from non-standard metrics and higher-curvature gravity \cite{tjoa2022unruhdewitt}. The UDW model and its covariant generalizations have been used to in various contexts in the field of relativistic quantum information (RQI) (see, e.g., \cite{pozas2015harvesting,pozas2016entanglement,tjoa2020harvesting,Tjoa2021notharvesting,smith2016topology,Aubry2018Vaidya,henderson2018harvesting,Gray2021imprint,Pipo2019without,Bruno2020neutrino,sahu2021sabotaging}, and references therein).

RQC protocols have been studied in various settings using UDW model as a base settings both perturbatively \cite{Cliche2010channel,Jonsson2017quantum} and nonperturbatively \cite{jonsson2018transmitting,Landulfo2016magnus1,Landulfo2021cost,Simidzija2020transmit,Koji2020superadditive}. 
However, even in the simplified UDW model,  nonperturbative methods in RQI are very rare, as is the case for anything involving quantum fields. There are essentially two nonperturbative approaches that have been in use in the literature. 
The first method involves assuming the detectors' free Hamiltonian to be \textit{gapless} \cite{Landulfo2016magnus1,Landulfo2021cost,Simidzija2018nogo}; the second method involves assuming the detectors to interact with the field very rapidly, effectively at one single instant in time, the so-called delta-coupling approach \cite{jonsson2018transmitting,Gallock2021nonperturbative,Simidzija2018nogo,Henderson2020Bandlimit,Simidzija2020transmit,Koji2020superadditive}. 
Both methods exploit the fact that in these two limits (gapless and delta-coupling) we can remove the time-ordering. The importance of nonperturbative methods cannot be overstated: as far as information-carrying capacity is concerned, perturbative methods define quantum channels that necessarily have perturbatively small (classical or quantum) channel capacity \cite{jonsson2018transmitting,Simidzija2020transmit}. 

In this work we will revisit the RQC between two UDW detectors interacting via a delta-coupling interaction, where each detector interacts locally with the field very rapidly (effectively instantaneous interaction at a single instant in time).
Our work sits in between \cite{Simidzija2020transmit} and \cite{Landulfo2016magnus1}, in that we are using the delta-coupling model discussed in \cite{Simidzija2020transmit} and analyze the classical channel capacity in the same spirit as \cite{Landulfo2016magnus1}. Our work is motivated by the fact that currently the classical channel capacity of delta-coupling model is not known (hence the \textit{entanglement-assisted} channel capacity\footnote{It was proven to be an entanglement breaking channel in \cite{jonsson2018transmitting,Simidzija2020transmit}, so \textit{unassisted} quantum channel capacity is zero.} is also unknown). More importantly, both delta coupling and gapless UDW models are conceptually similar in that they correspond to certain ``fast'' regime: for delta coupling, the interaction occurs much faster relative to the internal dynamics of the detectors and the field, while gapless regime corresponds to detectors' internal dynamics being much slower compared to the field's and the interaction timescale. However, they are generally catered to very different situations: for instance, gapless detectors can be used to study long-time phenomena such as thermalization \cite{Matsas2019rindler}, while delta coupling describes effectively instantaneous interaction at a single instant in time (in detector's frame).

In this work we show that as far as relativistic quantum communication are concerned, there is a sense in which relativistic communication via delta-coupled detectors and gapless detectors are equivalent. More precisely, we will show that in the delta-coupling UDW model, we can arrange and tune the detectors' parameters (such as energy gap, detector size, and interaction strength) such that the classical channel capacity is as good as the communication channel constructed nonperturbatively using gapless detectors in \cite{Landulfo2016magnus1}. Furthermore, we prove that this channel capacity is optimal: the maximum channel capacity for the delta-coupling model is given by the exact same algebraic expression, namely
\begin{align}
    C(\mathcal{E}) &= H\left(\frac{1}{2}+\frac{\nu_\textsc{b}}{2} \bigr|\cos(2\Delta(f_\textsc{a},f_\textsc{b}))\bigr|\right) - H\left(\frac{1}{2}+\frac{\nu_\textsc{b}}{2}\right)\,, \notag
\end{align}
where $H(x)$ is the Shannon entropy, $f_\textsc{a},f_\textsc{b}$ are the spacetime smearing (interaction region) of each detector, $\nu_\textsc{b}$ depends on Bob's spacetime smearing and the field state, and $\Delta(f_\textsc{a},f_\textsc{b})$ is the smeared field commutator\footnote{Note that $\Delta(f_\textsc{a},f_\textsc{b})$ is zero if both detectors are spacelike separated, so the channel capacity is zero, as it should be.}. In other words, the real difference between gapless detector model and delta-coupling model is ``in the details'' as far as RQC is concerned. This justifies the intuition that the two scenarios are in fact similar in relativistic communication settings. 

Our result comes with two small bonuses: first, it proves that with shared entanglement as a resource, the \textit{entanglement-assisted} quantum channel capacity, denoted $\mathcal{Q}_{\text{ea}}(\mathcal{E})$, for delta-coupling model can attain its minimum value of $1/2$  (in bits per unit time). This follows from the fact that $\mathcal{Q}_{\text{ea}}(\mathcal{E})\geq \frac{1}{2}C(\mathcal{E})$, and we will show that $C(\mathcal{E})$ can be made arbitrarily close to unity. Second, this work constitutes a generalization of delta-coupling UDW model to \textit{arbitrary} (globally hyperbolic) curved spacetimes, which is made possible by algebraic approach to quantum field theory instead of the more conventional canonical quantization approach. Since most delta-coupling calculations so far have been done in Minkowski space (see, e.g., \cite{Gallock2021nonperturbative,Henderson2020Bandlimit,Simidzija2018nogo,Simidzija2020transmit,Koji2020superadditive}), this generalization should prove useful for various future investigations.

Last but not least, this work aims partly to somewhat popularize a more algebraic language within (at least) the RQI community, hence our attempt to work in the language of algebraic quantum field theory (AQFT) in a way that hopefully makes the translation to the standard language using canonical quantization easier (see, e.g., \cite{wald1994quantum,Khavkhine2015AQFT,fewster2020algebraic}). Our convention and notation will be based on taking the best of \cite{Landulfo2016magnus1,fewster2020algebraic,Khavkhine2015AQFT} and \cite{Tales2020GRQO}.

Our paper is organized as follows. In Section~\ref{sec: AQFT} we introduce the algebraic framework for free scalar quantum field theory in curved spacetime. In Section~\ref{sec: UDW-model} we introduce the Unruh-DeWitt detector model and its delta-coupling variant where nonperturbative calculation is possible. In Section~\ref{sec: channel-capacity} we calculate the channel capacity of the communication channel between two detectors interacting very rapidly (delta-coupled) to the field. In Section~\ref{sec: results} we present our results and discussions. We adopt the units $c=\hbar=1$ and we use mostly-plus signature for the metric.

\section{Quantum field theory in curved spacetimes}
\label{sec: AQFT}

In this section we briefly review the algebraic formulation of quantum field theory for real scalar field. We hope that this can serve as a good pedagogical summary of the tools and helps popularizing the algebraic approach which has started to gain more usage in recent literature of RQI (see, e.g., \cite{Landulfo2016magnus1,Landulfo2021cost}).

\subsection{Algebra of observables}
Let us consider a free, real scalar field $\phi$ in (3+1)-dimensional globally hyperbolic Lorentzian spacetime $(\mathcal{M},g_{ab})$. The spacetime thus admits foliation by spacelike Cauchy surfaces $\Sigma_t$ labelled by real (time) parameter $t$. The field obeys the Klein-Gordon equation
\begin{align}
     P\phi = 0\,,\quad  P = -\nabla_a\nabla^a + m^2  + \xi R\,,
     \label{eq: KGE}
\end{align}
where $\xi \geq 0$, $R$ is the Ricci scalar and  $\nabla_a$ is the covariant derivative associated to Levi-Civita connection with respect to $g_{ab}$. 

Let $f\in \CS$ be a smooth compactly supported test function on $\M$, and  $E^\pm(\sx,\sy)$ be the retarded and advanced propagators associated to $P$ so that 
\begin{align}
    E^\pm f\equiv (E^\pm f)(\sx) \coloneqq \int \dd V'\, E^\pm (\sx,\sx')f(\sx') \,,
\end{align}
solves the inhomogeneous equation $P(E^\pm f) = f$. Here $\dd V' = \dd^4\sx'\sqrt{-g(\sx')}$ is the invariant volume element. The \textit{causal propagator} is defined by { $E=E^--E^+$. It can be shown that if $O$ is an open neighbourhood of some Cauchy surface $\Sigma$ and $\varphi$ is any real solution  to Eq.~\eqref{eq: KGE} with compact initial data, denoted $\varphi \in \Sol_\R(\M)$, then there exists $f\in \CS$ with $\text{supp}(f)\subset O$ such that $\varphi=Ef$ \cite{Khavkhine2015AQFT}. From this it follows that
\begin{align}
    P(Ef) = 0\,,\quad f\in \CS\,.
\end{align}
That is, $Ef$ is the solution to Klein-Gordon equation \eqref{eq: KGE}. }

In this work, we will adopt algebraic approach to quantum field theory (see Appendix~\ref{appendix: canonical-quantization} for comparison and see how it connects with canonical quantization approach). In algebraic quantum field theory (AQFT), field quantization is regarded as an $\R$-linear map from the space of smooth compactly supported test functions to a $*$-algebra $\A(\M)$
\begin{align}
    \phi: C^\infty_0(\mathcal{M})&\to \A(\M)\,,\quad f\mapsto \phi(f)\,,
\end{align}
which obeys the following conditions:
\begin{enumerate}[leftmargin=*,label=(\alph*)]
    \item (\textbf{Hermiticity}) $\phi(f)^\dag = \phi(f)$ for all $f\in \CS$;
    \item (\textbf{Klein-Gordon}) $\phi(Pf) = 0$ for all $f\in \CS$;
    \item (\textbf{Canonical commutation relations (CCR)}) $[\phi(f),\phi(g)] = -\ii\Delta(f,g)\openone $ for all $f,g\in \CS$, where
    \begin{align}
        {\Delta(f,g)\coloneqq -\int \dd V f(\sx) (Eg)(\sx)}\equiv -E(f,g)\,.
    \end{align}
    \item \textbf{(Time slice axiom)} Let $\Sigma\subset \M$ be a Cauchy surface and $O$ a fixed open neighbourhood of $\Sigma$. Then $\A(\M)$ is generated by the unit element $\openone$ and the smeared field operators $\phi(f)$ for all $f\in \CS$ with $\text{supp}(f)\subset O$.
\end{enumerate}
The $*$-algebra $\A(\M)$ is called the \textit{algebra of observables} of the real Klein-Gordon field\footnote{While conditions (a)-(c) are straightforward, condition (4) says that the algebra of observables do not contain other interesting physical observables such as the (smeared) stress-energy tensor $T_{\mu\nu}(f) \coloneqq \int \dd V \,T_{\mu\nu}(\sx)f(\sx)$; one needs to enlarge the algebra to incorporate these objects \cite{Khavkhine2015AQFT}.}. The smeared field operator reads
\begin{align}
    \phi(f) = \int \dd V\,\phi(\sx)f(\sx)\,,
\end{align}
hence in QFT $\phi(\sx)$ is to be interpreted as operator-valued distribution.

Once we have the algebra of observables, we need to provide a state. In AQFT, this is called an \textit{algebraic state}, defined by a $\C$-linear functional $\omega:\A(\M)\to \C$ such that 
\begin{align}
    \omega(\openone) = 1\,,\quad  \omega(A^\dagger A)\geq 0\quad \forall A\in \A(\M)\,.
    \label{eq: algebraic-state}
\end{align}
This is the algebraic statement that a quantum state is normalized to unity and gives non-negative expectation values for positive-semidefinite operators. The state $\omega$ is pure if it cannot be written as $\omega= \alpha \omega_1 + (1-\alpha)\omega_2$ for any $\alpha\in (0,1)$ and any two algebraic states $\omega_1,\omega_2$, i.e., $\omega=\omega_1=\omega_2$; otherwise we say the state is mixed. The Gelfand-Naimark-Segal (GNS) reconstruction theorem \cite{wald1994quantum,Khavkhine2015AQFT,fewster2020algebraic} then guarantees that we can construct a representation $\pi_\omega: \mathcal{\A(\M)}\to \mathcal{H}_\omega$ such that any algebraic state $\omega$ can be realized as a \textit{vector state} $\ket{\omega}$ in some Hilbert space $\mathcal{H}_\omega$, with the observables $A\in \A(\M)$ represented as operators  $\hat A\coloneqq \pi_\omega(A)$ acting on the Hilbert space. With GNS representation, the action of algebraic states take the familiar form
\begin{align}
    \omega(A) = \braket{\omega|\hat A|\omega}\,.
\end{align}
The main advantage of the AQFT approach is that it is independent of the representations of the CCR algebra chosen. Since quantum field theory in curved spacetimes admits infinitely many unitarily inequivalent representations of the CCR algebra, the algebraic framework allows us to deal with them all at once and only pick a representation by choosing specific state $\omega$.

Following \cite{Landulfo2016magnus1}, we will instead work with the ``exponentiated'' version of $\A(\M)$ called the \textit{Weyl algebra} (denoted by $\W(\M)$), because it is the elements of the Weyl algebra that appear directly in our calculation. The Weyl algebra $\W(\M)$ is a unital $C^*$-algebra generated by the elements
\begin{align}
    W(Ef) = e^{\ii\phi(f)}\,,\quad f\in \CS\,,
    \label{eq: Weyl-generator}
\end{align}
which satisfy \textit{Weyl relations}:
\begin{equation}
    \begin{aligned}
    W(Ef)^\dagger &= W(-Ef)\,,\\
    W(E (Pf) ) &= \openone\,,\\
    W(Ef)W(Eg) &= e^{\frac{\ii}{2}\Delta(f,g)} W(E(f+g))
    \end{aligned}
    \label{eq: Weyl-relations}
\end{equation}
where $f,g\in \CS$. The algebraic state with respect to this algebra is then the $\C$-linear functional $\omega:\W(\M)\to\C$ obeying \eqref{eq: algebraic-state} and one can again use the GNS representation theorem to obtain the Hilbert space formulation of the theory.

We remark in passing that at the level of the algebra of observables, despite the ``exponentiation'' in \eqref{eq: Weyl-generator} one \textit{cannot} interpret the smeared field operator $\phi(f)$ as the derivative $\partial_t\bigr|_{t=0}W(t Ef)$ since the algebra does not have the right topology \cite{fewster2020algebraic}; one should take the derivative of the GNS representation of the Weyl algebra observables instead: that is, if $\Pi_\omega:\W(\M)\to \mathcal{H}_\omega$ is a GNS representation with respect to $\omega$, then formally we do have 
\begin{align}
    \hat \phi(f)\coloneqq \pi_\omega(\phi(f)) = -\ii\frac{\dd}{\dd t}\Bigr|_{t=0}\Pi_\omega(e^{\ii t \phi(f)})\,,
\end{align}
where now $\hat\phi(f)$ is smeared field operator acting on Hilbert space $\mathcal{H}_\omega$. 

\subsection{Quasifree states}

One issue (or feature) with AQFT approach is that there are too many algebraic states available, not all of which are physically relevant. The consensus is that the physically relevant ones must be what is known as \textit{Hadamard states} \cite{Khavkhine2015AQFT}. Roughly speaking, these states have the right ``singular structure'' at short distances that respects local flatness property in general relativity\footnote{In other words, we require that if $f,g\in \CS$ have support that are very close to one another in $\M$, then quantities such as $\omega(W(Ef)W(Eg))$ must be very close to the corresponding value when $\M$ is Minkowski spacetime, see \cite{KayWald1991theorems} for more details.}. For our purposes, we are interested in Hadamard states that are also \textit{quasifree states}\footnote{These are also known as Gaussian states \cite{Khavkhine2015AQFT}, though some authors reserve ``Gaussian'' also for those that has non-vanishing odd-point correlators.}, denoted by $\omega_\mu$: these are the states which can be completely described by only its one-point correlators $\omega_\mu(\phi(f))$ and two-point correlators $\omega_\mu(\phi(f)\phi(g))$. Note that well-known field states such as the vacuum state, thermal state, and (squeezed) coherent states are all quasifree states, with thermal states (thermality defined according to Kubo-Martin-Schwinger (KMS) condition \cite{KayWald1991theorems}) being an example of mixed quasifree state.

The definition of quasifree state in AQFT given in the literature is somewhat convoluted because it relies on some optimization procedure; let us briefly clarify this. First, we note that classically the vector space of real-valued solutions of Klein-Gordon equation (with compact Cauchy data) $\Sol_\R(\M)$ can be made into a \textit{symplectic} vector space by equipping it with a symplectic form $\sigma:\Sol_\R(\M)\times\Sol_\R(\M)\to \R$:
\begin{align}
    \sigma(\varphi_1,\varphi_2) \coloneqq \int_{\Sigma_t}\dd\Sigma\, n^a\Bigr[\varphi_2\nabla_a\varphi_1 - \varphi_1\nabla_a\varphi_2\Bigr]\,,
\end{align}
where $n^a$ is unit normal to the Cauchy surface $\Sigma_t$ and this definition is independent of the Cauchy surface $\Sigma_t$. Any quasifree state $\omega_\mu$ is associated to a \textit{real inner product} $\mu:\Sol_\R(\M)\times \Sol_\R(\M)\to \R$ satisfying the inequality 
\begin{align}
    |\sigma(Ef,Eg)|^2\leq 4\mu(Ef,Ef)\mu(Eg,Eg)\,,
    \label{eq: real-inner-product}
\end{align}
for any $f,g\in \CS$. A quasifree state associated to $\mu$ is then defined by
\begin{align}
    \omega_\mu(W(Ef)) \coloneqq e^{-\mu(Ef,Ef)/2}\,.
\end{align}
However, as stated this expression is not helpful because it does not provide a way to calculate $\mu(Ef,Ef)$ from the inequality \eqref{eq: real-inner-product}.

In practice, the way to obtain the norm-squared $||Ef||^2 \coloneqq \mu(Ef,Ef)$ is to try to take a detour and make the space of solutions of the Klein-Gordon equation into a Hilbert space. It can be shown that we can always construct a Hilbert space $(\mathcal{H},\braket{\cdot,\cdot}_\mathcal{H})$ together with an $\R$-linear map $K:\Sol_\R(\M)\to \mathcal{H}$ such that \cite{KayWald1991theorems}
\begin{enumerate}[leftmargin=*,label=(\alph*)]
    \item $K\Sol_\R(\M)+\ii K\Sol_\R(\M)$ is dense in $\mathcal{H}$;
    \item $\mu(\varphi_1,\varphi_2)=\Re\braket{K\varphi_1,K\varphi_2}_\mathcal{H}$;
    \item {$\sigma(\varphi_1,\varphi_2) = - 2\Im\braket{K\varphi_1,K\varphi_2}_\mathcal{H}$.}
\end{enumerate}
The pair $(K,\mathcal{H})$ is called the \textit{one-particle structure} associated to the quasifree state $\omega_\mu$ \cite{KayWald1991theorems}.

In more familiar language, the linear map $K$ projects out the ``positive frequency part'' of real solution to the Klein-Gordon equation.
The smeared Wightman two-point function $\mathsf{W}(f,g)$ is then related to $\mu,\sigma$ by \cite{KayWald1991theorems}
\begin{align}
    \mathsf{W}(f,g) &= \omega(\phi(f)\phi(g)) \notag\\
    &= {\mu(Ef,Eg) - \frac{\ii}{2}\Delta(f,g)}\,.
\end{align}
Here we used the fact that $\sigma(Ef,Eg) = E(f,g) = -\Delta(f,g)$. Since $\Delta (f,f)=0$ due to antisymmetry of the CCR, we have that
\begin{align}
    ||Ef||^2 =  \mathsf{W}(f,f) = \braket{KEf,KEf}_\mathcal{H}\,.
    \label{eq: algebraic-norm}
\end{align}
Therefore, we can compute $\mu(Ef,Ef)$ if either (i) we know the Wightman two-point distribution of the theory associated to some quantum field state, or (ii) we know the inner product $\braket{\cdot,\cdot}_\mathcal{H}$ and how to project using $K$.

The inner product $\braket{\cdot,\cdot}_\mathcal{H}$ is precisely the \textit{Klein-Gordon inner product} $(\cdot,\cdot)_{\textsc{kg}}:\Sol_\C(\M)\times\Sol_\C(\M)\to \mathbb{C}$ restricted to $\mathcal{H}$, defined by extending $\sigma$ to complexified solutions, i.e.,
\begin{align}
    (\varphi_1,\varphi_2)_\textsc{kg} \coloneqq -\ii \sigma(\varphi_1^*,\varphi_2)\,,
    \label{eq: KG-inner-product}
\end{align}
where the symplectic form is now extended to \textit{complexified} solution $\Sol_\C(\M)$ of the Klein-Gordon equation. The restriction to $\mathcal{H}$ is required because $(\cdot,\cdot)_\textsc{kg}$ is not an inner product on $\Sol_\C(\M)$, as we will see later. In particular, we have
\begin{align}
    \Sol_\C(\M) \cong \mathcal{H}\oplus \overline{\mathcal{H}}\,,
\end{align}
where $\overline{\mathcal{H}}$ is the complex conjugate Hilbert space of $\mathcal{H}$ \cite{wald1994quantum}. It follows that the quasifree state can be written as
\begin{align}
    \omega_\mu(W(Ef)) = e^{-\frac{1}{2}\mathsf{W}(f,f)} = e^{-\frac{1}{2}||KEf||^2_{\textsc{kg}}}\,.
    \label{eq: norm-Ef}
\end{align}
This is the more useful expression that we wanted. 

Let us now make the connection to the usual (more practical) language of canonical quantization. We know that the Wightman two-point distribution is defined by
\begin{align}
    \mathsf{W}(\sx,\sy) &= \int \dd^3\bk\, u^{\phantom{*}}_\bk(\sx) u^*_\bk(\sy)\,,
\end{align}
where $u_\bk(\sx)$ are (positive-frequency) eigenfunctions of Klein-Gordon operator $P$ normalized with respect to Klein-Gordon inner product \eqref{eq: KG-inner-product}:
\begin{equation}
    \begin{aligned}
    (u_\bk,u_{\bk'})_\textsc{kg} &= \delta^3(\bk-\bk')\,,\quad (u_\bk,u^*_{\bk'})_\textsc{kg} = 0\,,\\
    (u_\bk^*,u^*_{\bk'})_\textsc{kg} &= -\delta^3(\bk-\bk')\,.
    \end{aligned}
    \label{eq: KG-normalization}
\end{equation}
Thus knowing the set of eigenmodes $\{u_\bk\}$ allows us to calculate the smeared Wightman function
\begin{align}
    \mathsf{W}(f,f) = \int \dd V\,\dd V' f(\sx)f(\sy)\mathsf{W}(\sx,\sy)\,.
    \label{eq: Wightman-double-smeared}
\end{align}
From the perspective of projection map $K$, what we are doing is projecting out the positive-frequency part of $Ef$ and express this in the positive-frequency basis $\{u_\bk\}$: that is, we have
\begin{align}
    Ef &= \int \dd^3\bk\, (u_\bk,Ef)_\textsc{kg} u_\bk + {(u_\bk,Ef)^*_\textsc{kg}u_\bk^*}\,,
\end{align}
so that using Eq.~\eqref{eq: KG-normalization} we get
\begin{align}
    KEf &= \int \dd^3\bk\, (u_\bk, Ef)_{\textsc{kg}} u_\bk(\sx)\,.
    \label{eq: KEf}
\end{align}
The restriction of the Klein-Gordon inner product to $\mathcal{H}$ gives
\begin{align}
    \braket{KEf,KEf}_\mathcal{H} &= (KEf,KEf)_\textsc{kg}\notag \\ 
    &= \int\dd^3\bk \,|(u_\bk,Ef)_\textsc{kg}|^2 \,.
    \label{eq: KG-norm-Ef}
\end{align}
By direct computation, we get
\begin{align}
    (u_\bk , Ef)_\textsc{kg} &= \ii \int\dd V \, u_\bk^*(\sx)f(\sx)\,,
    \label{eq: basis-Ef}
\end{align}
so that indeed we recover $\braket{KEf,KEf}_\mathcal{H} = \mathsf{W}(f,f)$. 

We close this section by remarking that if we wish to consider, for instance, the thermal KMS state, the procedure is very much similar except we need to pick a different algebraic state. This will in turn change the one-particle structure $(K,\mathcal{H})$ into $(K',\mathcal{H}')$ associated to the KMS state. It is in fact not hard to show that the only thing that changes in our calculation is the replacement of the $||Ef||^2$ in terms of the new one-particle structure: that is, for KMS state with KMS temperature $\beta$, we have \cite{KayWald1991theorems}
\begin{align}
    ||Ef||_{\beta}^2 
    &= \mathsf{W}_\beta(f,f)    
    \equiv \braket{K'Ef,K'Ef}_{\mathcal{H}'} \notag\\
    &= \braket{KEf,\coth(\beta\hat h/2)KEf}_{\mathcal{H}}\,,
\end{align}
where $W_\beta(f,f)$ is the smeared thermal Wightman distribution and $\hat h = \ii\partial_t$ is the ``one-particle Hamiltonian'' (the full Hamiltonian of the scalar field can be written as $\hat H_\phi = \openone\oplus\hat h\oplus (\hat h\otimes\hat h)+\cdots$ \cite{Landulfo2021cost}). This is another reason why algebraic approach is cleaner.

\section{Unruh-DeWitt model}
\label{sec: UDW-model}

In this work we start by writing the covariant generalization of the Unruh-DeWitt (UDW) detector model, which generalizes \cite{Unruh1979evaporation,DeWitt1979} for arbitrary spacetime smearing functions. We will simply call this the UDW model. We will first write down the general framework and then specialize to the case where the interaction is instantaneous.

\subsection{General Unruh-DeWitt model in curved spacetime}

Consider two observers Alice and Bob, each carrying a UDW detector. The UDW detector is taken to be a two-level quantum system (qubit) interacting locally with a quantized scalar field $\hat\phi$. The interaction is prescribed locally for each detector by a Hamiltonian density volume form {in the interaction picture}
\begin{align}
    \mathfrak{h}_{I,j} &= 
    \dd^n\sx\,\sqrt{-g}\,\left( f_j(\sx)\hat\mu_j(\tau_j(\sx))\otimes\hat\phi(\sx)\right)\,,
\end{align}
where $f_j$ is spacetime smearing\footnote{Note that in order to match the language in AQFT, the coupling constants typically used in UDW literature that prescribes the coupling strength of the detector-field system  (denoted by $\lambda_j$) is absorbed into the definition of $f_j$, see Section~\ref{subsec: example} for an explicit example.} of detector $j$. The monopole operator $\hat\mu_j(\tau_j)$ is obtained by evolving Pauli-$X$ operator $\hat \sigma_j^x$ via each detector's free Hamiltonian $\hat H_{\textsc{d},j} = \frac{\Omega_j}{2}\hat\sigma^+_j\hat \sigma^-_j = \frac{\Omega_j}{2}(\hat\sigma^z_j+\openone)$: 
\begin{align}
    \hat\mu_j(\tau_j) 
    &= \hat\sigma^+_j e^{\ii\Omega_j\tau_j} + \hat\sigma^-_j e^{-\ii\Omega_j\tau_j}\,,
\end{align}
where $\Omega_j$ and $\tau_j$ are the energy gap and the proper time of detector $j$. 

Since we have a communication protocol involving two detectors, the unitary time evolution is given by time-ordered exponential
\begin{align}
    \hat U = \mathcal{T}\exp\left[-\ii\int_\mathcal{M}\mathfrak{h}_{I,\textsc{a}}+\mathfrak{h}_{I,\textsc{b}}\right]\,.
\end{align}
The quantum channel $\mathcal{E}: \mathcal{D}(\mathcal{H}_\textsc{a})\to\mathcal{D}(\mathcal{H}_\textsc{b})$ between the spaces of density matrices $\mathcal{D}(\mathcal{H}_\textsc{a}),\mathcal{D}(\mathcal{H}_\textsc{b})$ defining the relativistic communication protocol is naturally given in the Stinespring representation by
\begin{align}
    \rho_\textsc{b}\coloneqq \mathcal{E}(\rho_{\textsc{a},0}) = \tr_{\textsc{a},\phi}\left[\hat U(\rho_{\textsc{a,0}}\otimes \rho_{\textsc{b,0}}\otimes\rho_{\phi})\hat U^\dagger\right]\,,
    \label{eq: channel}
\end{align}
where $\rho_{j,0}$ is the initial states of detector $j$ and $\rho_\phi$ is the density matrix of the quantum field. After expanding the unitary operator on total system, the partial trace over the field's degrees of freedom can be written more covariantly if we use the fact that\footnote{We could also do the same for the qubits if we are willing to do ``algebraic'' quantum information, but we will not do it here since it does not serve much purpose in this work.}
\begin{align}
    \omega(A) = \tr_\phi(\rho_\phi\hat A)\,,
\end{align}
using the fact that $\omega(A)$ evaluates the expectation value of an operator $A\in \mathcal{W}(\mathcal{M})$ with respect to the algebraic state $\omega$ of the field. We write $\omega$ instead of $\omega_\mu$ for convenience since we are going to consider only one algebraic state in this work, namely that the vacuum state of its Fock representation.

This is as far as one gets with the general framework without specifying the spacetime $(\mathcal{M},g)$ and the spacetime smearing functions $f_j$ which prescribe where in spacetime the interaction between each detector and the field takes place, as well as the algebraic state $\omega$.

\subsection{Quantum channel for delta interaction}

The delta interaction is an approximation where the detector interacts with the field at very short timescale that it can be approximated as occurring at one instant of time in each detector's frame. However, this requires us to be able to specify a notion of one instant of time in each detector's frame. A natural way to do this is to consider the Fermi normal coordinates (FNC) of each detector, labelled by coordinates $\mathsf{z} = (\tau,\overline{\bm{x}})$, where $\sz(\tau)=(\tau,\overline{\bm{x}}=\mathbf{0})$ is the centre of mass coordinates of the detector with proper time $\tau$ \cite{Tales2020GRQO}.

For this to work, we assume that the qubit detector is \textit{rigid} in its own frame, so that we can write in FNC
\begin{align}
    f_j(\sz) = \lambda_j \chi_j(\tau)F_j(\overline{\bm{x}})\,,
    \label{eq: FNC-factorization}
\end{align}
where $\chi_j$ is the \textit{switching function} of the detector prescribing the duration of interaction, and $F_j(\overline{\bm{x}})$ is the spatial profile of the detector giving the effective size of the detector. Note that Eq.~\eqref{eq: FNC-factorization} is the statement that the spatial profile $F_j(\overline{\bm{x}})$ is effectively unchanged throughout the interaction region, hence the qubit is \textit{rigid} in shape (in the sense of Born rigidity)\footnote{This is not going to be true in, say, detectors undergoing very large acceleration, or spacetimes with extremely high curvature. Born rigidity says that this approximation holds when $a L/c^2\ll 1$, where $c$ is the speed of light and $L$ is the approximate size/diameter of the detector. So for all practical purposes, this will be very reasonable approximation.}. Under this factorization, we can now implement delta switching by setting
\begin{align}
    \chi_j(\tau_j) = \eta_j\delta(\tau_j-\tau_{j,0})\,,
    \label{eq: switching}
\end{align}
where $\eta_j$ has units of time to maintain $\chi_j$ dimensionless and $\tau_{j,0}$ denotes the switch-on time of detector $j$. For simplicity we set $\eta_j = \eta>0$ for both detectors.

The delta interaction allows for nonperturbative calculation because the full unitary can now be written as a simple product of exponentials instead of time-ordered one:
\begin{align}
    \hat U &= \hat U_\textsc{b} \hat U_\textsc{a}\,,\quad \hat U_j = \exp\left[ -\ii \hat\mu_j(\tau_{j,0}) \otimes \hat Y_j \right]\,,
\end{align}
where the smeared field operator $\hat Y_j$ is given by
\begin{align}
    \hat Y_j 
    &\coloneqq \Tilde{\lambda}_j \int_{\tau = \tau_{j,0}}\!\!\!\dd^3\overline\bx \sqrt{-g}\,F_j(\overline\bx)\hat\phi(\overline\bx)\,,
\end{align}
where $\tilde\lambda_j \coloneqq \lambda_j\eta$. The unitary $\hat U_j$ then reads
\begin{align}
    \hat U_j = \openone_j\otimes \cos \hat{Y}_j -\ii\hat\mu_j(\tau_{j,0})\otimes \sin\hat{Y}_j  \,.
    \label{eq: Uj}
\end{align}

Substituting Eq.~\eqref{eq: Uj} into Eq.~\eqref{eq: channel}, we get
\begin{align}
    \rho_\textsc{b} 
    &=\mathcal{E}(\rho_{\textsc{a,0}})  \notag\\
    &= (\gamma_{cccc}+\gamma_{cssc})\rho_\textsc{b,0}\, + \notag\\
    &\hspace{0.5cm} (\gamma_{sccs}+\gamma_{ssss}) \hat\mu(\tau_{\textsc{b,0}})\rho_\textsc{b,0}\hat\mu(\tau_{\textsc{b,0}}) \,+ \notag\\
    &\hspace{0.5cm} (\gamma_{scsc}-\gamma_{sscc}){\theta(\tau\ts{a,0})}[\hat\mu(\tau_\textsc{b,0}),\rho_\textsc{b,0}]\,,
    \label{eq: final-state-Bob}
\end{align}
{where we define $\theta(\tau\ts{a,0})\coloneqq  \tr[ \rho\ts{a,0} \hat \mu\ts{a}(\tau\ts{a,0}) ]$} and the constants $\gamma_{ijkl}$ are defined by
\begin{align}
    \gamma_{ijkl} 
    &\coloneqq 
    \tr (\hat X_{\textsc{b}}^{(i)} \hat X_{\textsc{a}}^{(j)}
    \rho_\phi 
    \hat X_{\textsc{a}}^{(k)}\hat X_{\textsc{b}}^{(l)}) \,,  \\
    \hat X_{\alpha}^{(c)}&= \cos\hat Y_{\alpha}\,,\quad \hat X^{(s)}_{\alpha} = \sin\hat Y_{\alpha}\,, 
\end{align}
with $\rho_\phi$ the initial density matrix of the field associated to algebraic state $\omega$. The constants $\gamma_{ijkl}$ can be written in terms of $\omega$ by slight reordering:
\begin{align}
    \gamma_{ijkl} 
    &= \omega( \hat X_{\textsc{a}}^{(k)}\hat X_{\textsc{b}}^{(l)}
    \hat X_{\textsc{b}}^{(i)} 
    \hat X_{\textsc{a}}^{(j)}) \\
    &\equiv \tr (\rho_\phi \hat X_{\textsc{a}}^{(k)}
    \hat X_{\textsc{b}}^{(l)}
    \hat X_{\textsc{b}}^{(i)} \hat X_{\textsc{a}}^{(j)}) \notag \,.
\end{align}
The explicit expressions for each $\gamma_{ijkl}$ are given in Appendix~\ref{appendix: gammas}: for the quasifree state $\omega$ we consider here, the expressions simplify greatly\footnote{This is because for quasifree states the ``one-point functions'' vanish, i.e., $\omega(\phi(f))=0$. However, the general formalism considered here will carry over for arbitrary Gaussian states with  non-vanishing one-point functions $\omega(\phi(f))\geq 0$.} so that the constant prefactors in Eq.~\eqref{eq: final-state-Bob} now reads
\begin{subequations}
    \begin{align}
        \gamma_{cccc}+\gamma_{cssc} &= \frac{1}{2}+\frac{\nu_\textsc{b}}{2}\cos(2
        \Delta(f_\textsc{a},f_\textsc{b})) \,,\\
        \gamma_{sccs}+\gamma_{ssss} &= \frac{1}{2}-\frac{\nu_\textsc{b}}{2}\cos(2
        \Delta(f_\textsc{a},f_\textsc{b}))\,,\\
        \gamma_{scsc} - \gamma_{sscc} &= -\frac{\ii \nu_\textsc{b}}{2}\sin(2\Delta(f_\textsc{a},f_\textsc{b}))\,,
    \end{align}
\end{subequations}
where
\begin{align}
    \nu_\textsc{b} &= \omega(W(2Ef_\textsc{b})) =  e^{-2||Ef_\textsc{b}||^2} = e^{-2\mathsf{W}(f_\textsc{b},f_\textsc{b})}\,. 
    \label{eq: nub-main}
\end{align}
Eq.~\eqref{eq: final-state-Bob} obtained here turns out to be structurally very similar to Eq.~(45) in \cite{Landulfo2016magnus1} for gapless detectors. It is also worth emphasizing that the expressions for $\gamma_{ijkl}$ above and computed in Appendix~\ref{appendix: gammas} are very general as they only employ the Weyl relations, thus valid for arbitrary curved spacetimes.

\section{Channel capacity of the quantum channel}
\label{sec: channel-capacity}

In this section we briefly review the concept of classical and quantum channel capacity. We will calculate the channel capacity for the quantum channel $\mathcal{E}$ and obtain an optimal bound for the channel capacity. We will then give a concrete example in Minkowski spacetime and compare our results with the channel obtained using gapless detector model in \cite{Landulfo2016magnus1}.

\subsection{Classical and quantum channel capacity}

{Suppose that Alice wishes to transmit a message chosen from the set of messages $\{1,2,...,N\}$ using the quantum channel $\mathcal{E}$. One way to do this is to choose a block coding scheme and then apply $n$ independent uses of the channel. The coding scheme is as follows. Let $X$ be a random variable with $N$ outcomes corresponding to Alice's choice of messages. For each $m\in X$, Alice prepares a (possibly entangled) state  $\rho_{\textsc{A}0,m}^{(n)}\in  \mathcal{D}(\mathcal{H}^{\otimes n}_\textsc{a})$ as input state of the $n$-independent uses of the channel. That is, Alice applies $\mathcal{E}^{\otimes n}$ acting on $\rho_{\textsc{A}0,m}^{(n)}$ to transmit the message to Bob.} 

On Bob's side, he receives the state
\begin{align}
    \rho_{\textsc{b},m}^{(n)}
    \coloneqq \mathcal{E}^{\otimes n}(\rho_{\textsc{a0},m}^{(n)}) \in \mathcal{D}(\mathcal{H}_\textsc{b}^{\otimes n})\,.
\end{align}
Bob will decode the message by choosing a suitable POVM $\{\hat E_{m}: m\in Y\}$ acting on $\rho^{(n)}_{\textsc{b},m}$, where $Y=\{1,2,..., N'\}$ with $|Y|\geq |X|$. The probability he decodes the message correctly is given by the conditional probability
\begin{align}
    \Pr(Y=m|X=m) &= \tr\left(\hat E_m^{\phantom{\otimes}} \rho_{\textsc{b},m}^{(n)}\right)\,,
\end{align}
with probability of error $1-\Pr(Y=m|X=m)$. 

Now we define the total initial state of Alice $\rho_{\textsc{a},0}$ as an ensemble $\{p_m, \rho_{\textsc{a0},m}\}$ given by
\begin{align}
    \rho_{\textsc{a},0} \coloneqq \sum_{m}p_m \rho_{\textsc{a}0,m}
    \label{eq: ensemble-Alice}
\end{align}
where $\sum_m p_m=1$ and $p_m\geq 0$ are the probabilities of the ensemble. 

The \textit{Holevo $\chi$ quantity} of an ensemble $\rho = \sum_{m}p_m\rho_m$ is defined as
\begin{align}
    \chi(\rho) \coloneqq S(\rho) - \sum_{m}p_mS(\rho_{m})\,,
\end{align}
where $S(\rho) = -\tr(\rho\log_2\rho)$ is the von-Neumann entropy (or $S(\rho) = -\sum_j\lambda_j\log_2\lambda_j$ where $\lambda_j$ is the positive eigenvalues of $\rho$). The Holevo information of a channel $\mathcal{E}$ is then given by the maximization of Holevo $\chi$ quantity $\chi(\mathcal{E}(\rho))$ over all possible ensembles $\rho=\sum_{m}p_m \rho_{m}$:
\begin{align}
    \chi(\mathcal{E})\coloneqq \max_{\{p_m,\rho_m\}}\chi(\mathcal{E}(\rho))\,. \label{eq:Holevo info}
\end{align}
The Holevo-Schumacher-Westmoreland (HSW) theorem then states that the classical channel capacity $C(\mathcal{E})$ of a quantum channel $\mathcal{E}$ can be written as an asymptotic limit \cite{Wilde2013textbook,nielsen2000quantum}
\begin{align}
    C(\mathcal{E}) = \lim_{n\to\infty}\frac{1}{n}\chi(\mathcal{E}^{\otimes n})\,.
\end{align}

The channel capacity $C(\mathcal{E})$ is in general difficult to calculate because of the asymptotic limit of large number of $n$ independent uses of the channel. However, it can be shown that the channel in Eq.~\eqref{eq: final-state-Bob} is in fact \textit{entanglement breaking channel} \cite{Simidzija2020transmit}. The fact that it is entanglement-breaking implies that \cite{Shor2002entanglementbreak} 
\begin{align}
    \chi(\mathcal{E}^{\otimes n}) = n\chi(\mathcal{E})\,,
\end{align}
so that the channel capacity for our channel reduces to
\begin{align}
    C(\mathcal{E})= \chi(\mathcal{E})\,.\label{eq:classical capacity def}
\end{align}
This is a much more tractable expression since we only need to maximize over the ensembles without taking any asymptotic limits.

Note that since $\mathcal{E}$ is entanglement-breaking, it automatically implies that the \textit{quantum} channel capacity is zero (see, e.g., \cite{Simidzija2020transmit,Simidzija2018nogo} and \cite{holevo2008entanglement}). Therefore, we cannot use this channel as-is to transmit genuine quantum information. However, as we will see in the next subsection, the classical channel capacity can be made arbitrarily close to unity with suitable choice of parameters. This means that if both Alice and Bob have \textit{shared entanglement}, they can have quantum channel capacity of at least $\frac{1}{2}C(\mathcal{E})$ \cite{Wilde2013textbook}, so that this channel is not completely useless for transmission of quantum information.

\subsection{Maximization of channel capacity $C(\mathcal{E})$}
Let us determine the classical channel capacity $C(\mathcal{E})$ of the quantum channel $\mathcal{E}$. 
To this end, we will obtain the eigenvalues of $\mathcal{E}(\rho\ts{a0,$m$})$ and $\mathcal{E}(\rho\ts{a,0})$, and maximize the Holevo $\chi$ quantity $\chi(\mathcal{E}(\rho\ts{a,0}))$ over an ensemble $\{ p_m, \rho\ts{a0,$m$} \}$ at the end. 

Since Alice has a qubit system, each $\rho\ts{a0,$m$}$ can be written in terms of eigenbasis $\{\ket{0},\ket{1}\}$ of $\hat \sigma_z$:
\begin{align}
    \rho\ts{a0,$m$}
    &=
        \dfrac{1}{2} 
        (\id\ts{a} + \bm{r}_m \cdot \bm{\sigma}\ts{a}) \\
    &=
        \dfrac{1+z_m}{2} \ket{0\ts{a}} \bra{0\ts{a}}
        +
        \dfrac{ x_m - \ii y_m }{2} \ket{0\ts{a}}\bra{1\ts{a}} \notag \\
        &\hspace{0mm}
        +
        \dfrac{ x_m + \ii y_m }{2} \ket{1\ts{a}}\bra{0\ts{a}}
        +
        \dfrac{1-z_m}{2} \ket{1\ts{a}} \bra{1\ts{a}}\,,
        \label{eq: Alice-m-ensemble}
\end{align}
where $\bm{r}_m \coloneqq (x_m, y_m, z_m)$ is the Bloch vector satisfying $x_m^2 + y_m^2 + z_m^2 \leq 1$ and $\bm{\sigma}\ts{a}\coloneqq (\hat \sigma\ts{a}^x, \hat \sigma\ts{a}^y, \hat \sigma\ts{a}^z)$. 
Then we have 
\begin{align}
    \theta_m(\tau\ts{a,0})
    &\coloneqq
        \tr[ \rho\ts{a0,$m$} \hat \mu\ts{a}(\tau\ts{a,0}) ] \notag \\
    &=
        x_m \cos (\Omega\ts{a} \tau\ts{a,0})
        +
        y_m \sin (\Omega\ts{a} \tau\ts{a,0})\,. \label{eq:thetaAlice}
\end{align}
Similarly, Bob's initial state $\rho\ts{b,0}$ can be written as 
\begin{align}
    \rho\ts{b,0}
    &=
        \dfrac{1+z\ts{b}}{2} \ket{0\ts{b}} \bra{0\ts{b}}
        +
        \dfrac{ x\ts{b} - \ii y\ts{b} }{2} \ket{0\ts{b}}\bra{1\ts{b}} \notag \\
        &\hspace{0mm}
        +
        \dfrac{ x\ts{b} + \ii y\ts{b} }{2} \ket{1\ts{b}}\bra{0\ts{b}}
        +
        \dfrac{1-z\ts{b} }{2} \ket{1\ts{b}} \bra{1\ts{b}}\,, 
\end{align}
with $x\ts{b}^2 + y\ts{b}^2 + z\ts{b}^2 \leq 1$. 

Evaluating the action of quantum channel $\mathcal{E}$ defined in Eq.~\eqref{eq: final-state-Bob} on Alice's collection of states $\rho_{\textsc{a}0,m}$ in Eq.~\eqref{eq: Alice-m-ensemble}, we get 
\begin{widetext}
\begin{align}
    \mathcal{E}( \rho\ts{a0,$m$} )
    &=  r_{11}^m \ket{0\ts{b}}\bra{0\ts{b}} +  
        r_{12}^m \ket{0\ts{b}} \bra{1\ts{b}}
       +r_{12}^{m *} \ket{1\ts{b}} \bra{0\ts{b}} +
        r_{22}^m \ket{1\ts{b}} \bra{1\ts{b}}\,,
\end{align}
where 
\begin{subequations}
\begin{align}
    r_{11}^m
    &=
        \dfrac{1}{2}
        \Bigkako{
            1 + \nu_\textsc{b} z\ts{b} \cos(2\Delta(f\ts{a}, f\ts{b}))
            + 
            \nu\ts{b} \theta_m(\tau\ts{a,0}) \sin(2\Delta(f\ts{a}, f\ts{b}))
            \Bigkagikako{
                y\ts{b} \cos (\Omega\ts{b} \tau\ts{b,0} )
                +
                x\ts{b} \sin (\Omega\ts{b} \tau\ts{b,0} )
            }
        }, \\
    r_{12}^m
    &=
        \dfrac{1}{4}
        \Bigkako{
            e^{ 2\ii \Omega\ts{b}\tau\ts{b,0} }
            (x\ts{b} + \ii y\ts{b}) 
            \Big[ 1-\nu\ts{b} \cos (2\Delta(f\ts{a}, f\ts{b}))\Big]
            +
            (x\ts{b} - \ii y\ts{b}) 
            \Big[ 1+\nu\ts{b} \cos (2\Delta(f\ts{a}, f\ts{b})) \Big] \notag \\
            &\hspace{1cm}
            + 
            2\ii e^{ \ii \Omega\ts{b} \tau\ts{b,0} }
            \nu\ts{b} z\ts{b} \theta_m(\tau\ts{a,0}) \sin(2\Delta(f\ts{a}, f\ts{b}))
        }, \\
    r_{22}^m
    &=
        \dfrac{1}{2}
        \Bigkako{
            1 - \nu\ts{b} z\ts{b} \cos(2\Delta(f\ts{a}, f\ts{b}))
            - 
            \nu\ts{b} \theta_m(\tau\ts{a,0}) \sin(2\Delta(f\ts{a}, f\ts{b}))
            \Bigkagikako{
                y\ts{b} \cos (\Omega\ts{b} \tau\ts{b,0} )
                +
                x\ts{b} \sin (\Omega\ts{b} \tau\ts{b,0} )
            }
        }. 
\end{align}
\end{subequations}
Next, evaluating the action of quantum channel $\mathcal{E}$  on Alice's total state $\rho_{\textsc{a},0}$ in Eq.~\eqref{eq: ensemble-Alice}, we get 
\begin{align}
    \mathcal{E}( \rho\ts{a,0} )
    &=
        \mathcal{E}
        \kako{
            \sum_m p_m \rho\ts{a0,$m$}
        } =
        r_{11} \ket{0\ts{b}}\bra{0\ts{b}}
        +
        r_{12} \ket{0\ts{b}} \bra{1\ts{b}}
          +
        r_{12}^{*} \ket{1\ts{b}} \bra{0\ts{b}}
        +
        r_{22} \ket{1\ts{b}} \bra{1\ts{b}}\,,
\end{align}
where 
\begin{subequations}
\begin{align}
    r_{11}
    &=
        \dfrac{1}{2}
        \Bigkako{
            1 + \nu\ts{b} z\ts{b} \cos(2\Delta(f\ts{a}, f\ts{b}))
            + 
            \nu\ts{b} \theta(\tau\ts{a,0}) \sin(2\Delta(f\ts{a}, f\ts{b}))
            \Bigkagikako{
                y\ts{b} \cos (\Omega\ts{b} \tau\ts{b,0} )
                +
                x\ts{b} \sin (\Omega\ts{b} \tau\ts{b,0} )
            }
        }, \\
    r_{12}
    &=
        \dfrac{1}{4}
        \Bigkako{
            e^{ 2\ii \Omega\ts{b} \tau\ts{b,0} }
            (x\ts{b} + \ii y\ts{b}) 
            \Big[ 1-\nu\ts{b} \cos (2\Delta(f\ts{a}, f\ts{b})) \Big]
            +
            (x\ts{b} - \ii y\ts{b}) 
            \Big[ 1+\nu\ts{b} \cos (2\Delta(f\ts{a}, f\ts{b}))\Big] \notag \\
            &\hspace{1cm}
            + 
            2\ii e^{ \ii \Omega\ts{b} \tau\ts{b,0} }
            \nu\ts{b} z\ts{b} \theta(\tau\ts{a,0}) \sin(2\Delta(f\ts{a}, f\ts{b}))
        }, \\
    r_{22}
    &=
        \dfrac{1}{2}
        \Bigkako{
            1 - \nu\ts{b} z\ts{b} \cos(2\Delta(f\ts{a}, f\ts{b}))
            - 
            \nu\ts{b} \theta(\tau\ts{a,0}) \sin(2\Delta(f\ts{a}, f\ts{b}))
            \Bigkagikako{
                y\ts{b} \cos (\Omega\ts{b} \tau\ts{b,0} )
                +
                x\ts{b} \sin (\Omega\ts{b} \tau\ts{b,0} )
            }
        }\,,
\end{align}
\end{subequations}
\end{widetext}
with
\begin{align}
    \theta(\tau\ts{a,0})
    &\equiv
        \sum_m p_m \theta_m(\tau\ts{a,0}). 
\end{align}

In order to calculate the channel capacity, we will need the eigenvalues of $\rho_\textsc{b}=\mathcal{E}(\rho_\textsc{a,0})$ as well as eigenvalues of $\mathcal{E}(\rho\ts{a0,$m$})$.  Let $p_{\mathcal{E}_m}^\pm$ and $p_{\mathcal{E}}^\pm$ be the eigenvalues of $\mathcal{E}(\rho\ts{a0,$m$})$ and $\mathcal{E}(\rho\ts{a,0})$ respectively. They are given by 
\begin{align}
    p_{\mathcal{E}_m}^\pm
    &=
        \dfrac{1}{2} 
        \pm\frac{1}{2}
        \sqrt{ 
            \mathcal{P}^2
            + 
            \nu\ts{b}^2 \mathcal{R}_m
        }, \\
    p_{\mathcal{E}}^\pm
    &=
        \dfrac{1}{2} 
        \pm\frac{1}{2}
        \sqrt{ 
            \mathcal{P}^2
            + 
            \nu\ts{b}^2 \mathcal{R}
        }, 
\end{align}
where
\begin{subequations}
\begin{align}
    \mathcal{P}
    &\equiv
       x\ts{b}\cos(\Omega\ts{b} \tau\ts{b,0}) - y\ts{b}\sin(\Omega\ts{b} \tau\ts{b,0}) \\
    \mathcal{R}_m
    &\equiv
        \Bigkagikako{
            \theta_m^2(\tau\ts{a,0}) \sin^2(2\Delta(f\ts{a}, f\ts{b})) + \cos^2(2\Delta(f\ts{a}, f\ts{b}))
        } \notag \\
        &\hspace{0.5cm}
        \times 
        \kako{
            r\ts{b}^2
            -\mathcal{P}^2
        }, \\
    \mathcal{R}
    &\equiv
        \Bigkagikako{
            \theta^2(\tau\ts{a,0}) \sin^2(2\Delta(f\ts{a}, f\ts{b})) + \cos^2(2\Delta(f\ts{a}, f\ts{b}))
        }\notag \\
        &\hspace{0.5cm}
        \times 
        \kako{
            r\ts{b}^2
            -\mathcal{P}^2
        },
\end{align}
\end{subequations}
and $r\ts{b}\equiv \sqrt{ x\ts{b}^2 + y\ts{b}^2 + z\ts{b}^2 }$ is the length of Bloch vector for Bob's initial state $\rho_\textsc{b,0}$.

Now we claim that our channel capacity $C(\mathcal{E})$ functionally reduces to the one in \cite{Landulfo2016magnus1} when $\mathcal{P}=0$. 
To see this, suppose that Bob prepares his initial state as $(x\ts{b}, y\ts{b}, z\ts{b})=(0,0,1)$. Then the eigenvalues $p_{\mathcal{E}_m}^+$ and $p_{\mathcal{E}}^+$ will have the following bounds:
\begin{align}
    p_{\mathcal{E}_m}^+
    &\leq 
        \dfrac{1}{2} 
        +
        \dfrac{\nu\ts{b}}{2}, \\
    p_{\mathcal{E}}^+
    &\geq 
        \dfrac{1}{2} 
        + \dfrac{\nu\ts{b}}{2}|\cos (2\Delta(f\ts{a},f\ts{b}))| .
\end{align}
Here, we used the fact that $0\leq \theta_m^2(\tau\ts{a,0})\leq 1$. 
Since $\nu\ts{b}\geq 0$ and the von Neumann entropy is monotonically decreasing for $p_{\mathcal{E}_m}^+, p_{\mathcal{E}}^+ > 1/2$, we obtain a bound for the Holevo $\chi$ quantity $\chi(\mathcal{E}(\rho\ts{a,0}))$:
\begin{align}
    &\chi(\mathcal{E}(\rho\ts{a,0}))
    \leq \notag \\
        &H
        \kako{
            \dfrac{1}{2} 
            + 
            \dfrac{\nu\ts{b}}{2} |\cos (2\Delta(f\ts{a},f\ts{b}))|
        }
        -
        H
        \kako{
            \dfrac{1}{2} + \dfrac{\nu\ts{b}}{2}
        },
\end{align}
where $H(x)$ is the Shannon entropy. 
This bound is in fact functionally identical to the one calculated in \cite{Landulfo2016magnus1} for gapless detectors. 
In our case, the equality can be achieved if Alice prepares her initial state to be $\rho_{\textsc{a,0}}=\sum_{m}p_m\rho_{\textsc{a0},m}$ where $p_1=p_2=1/2$, $p_m=0$ for $m\geq 3$, with $\rho_{\textsc{a0},1}$ associated to Bloch vector $\bm{r}_1=(\cos(\Omega\ts{A}\tau\ts{A,0}), \sin(\Omega\ts{A}\tau\ts{A,0}),0)$, and $\rho_{\textsc{a0},2}$ associated to Bloch vector 
$\bm{r}_2=(-\cos(\Omega\ts{A}\tau\ts{A,0}), -\sin(\Omega\ts{A}\tau\ts{A,0}),0)$. 
With this choice we have $\theta_m^2(\tau\ts{A,0})=1$ for $m=1,2$ (and zero for $m\geq 3$) and $\theta(\tau\ts{A,0})=0$.  Then from \eqref{eq:Holevo info} and \eqref{eq:classical capacity def}, we obtain the classical channel capacity 
\begin{align}
    C(\mathcal{E})
    &=
        H
        \kako{
            \dfrac{1}{2} 
            + 
            \dfrac{\nu\ts{b}}{2} |\cos (2\Delta(f\ts{a},f\ts{b}))|
        }
        -
        H
        \kako{
            \dfrac{1}{2} + \dfrac{\nu\ts{b}}{2}
        }. \label{eq:classical capacity}
\end{align}
Note that Bob's phase $\Omega\ts{b}\tau\ts{b,0}$ is arbitrary; \eqref{eq:classical capacity} holds for any $\Omega\ts{b}\tau\ts{b,0}$ as long as Bob initially prepares $(x\ts{b}, y\ts{b}, z\ts{b})=(0,0,1)$. Observe that the channel $\mathcal{E}$ depends on detectors' energy gap via the eigenvalues $p^\pm_{\mathcal{E}}$ and $p^\pm_{\mathcal{E}_m}$, but the optimal channel capacity $C(\mathcal{E})$ in Eq.~\eqref{eq:classical capacity} is independent of the energy gaps.

In fact we can obtain the channel capacity in a form very close to Eq.~\eqref{eq:classical capacity} without specifying Bob's initial state, by instead tuning Bob's phase $\Omega\ts{b} \tau\ts{B,0}$.  
Let us choose $\Omega\ts{b} \tau\ts{B,0}$ to be 
\begin{align}
    \Omega\ts{b} \tau\ts{b,0} + \alpha = n\pi,~~~(n \in \mathbb{Z})
\end{align}
where 
\begin{align}
    \cos \alpha 
    &=
        \dfrac{ -y\ts{b} }{ \sqrt{ x\ts{b}^2 + y\ts{b}^2 } },~
    \sin \alpha 
    =
        \dfrac{ x\ts{b} }{ \sqrt{ x\ts{b}^2 + y\ts{b}^2 } },
\end{align}
so that $\mathcal{P}=0$. 
Then the eigenvalues $p_{\mathcal{E}_m}^+$ and $p_{\mathcal{E}}^+$ will be bounded by 
\begin{align}
    p_{ \mathcal{E}_m }^+
    &\leq
        \dfrac{1}{2} + \dfrac{\nu\ts{b} r\ts{b}}{2}, \\
    p_{\mathcal{E}}^+
    &\geq 
        \dfrac{1}{2} 
        +
        \dfrac{\nu\ts{b} r\ts{b}}{ 2} 
        |\cos (2\Delta (f\ts{a}, f\ts{b}))|\,,
\end{align}
and the channel capacity takes the same form as Eq.~\eqref{eq:classical capacity} but modified as $\nu\ts{b}\to \nu\ts{b} r\ts{b}$.  It tells us that one can obtain a similar form of \eqref{eq:classical capacity} by tuning Bob's detector, and $C(\mathcal{E})$ depends on the length $r\ts{b}$ of Bob's initial Bloch vector. It is straightforward now to see that such a capacity $C(\mathcal{E})$ achieves the maximum if Bob prepares a pure state $r\ts{b}=1$, but channel capacity gets worse if he uses initially maximally mixed state $r\ts{b}=0$; therefore, the channel capacity in Eq.~\eqref{eq:classical capacity} is optimal for $r_\textsc{b}=1$.

The main result of this work is the channel capacity $C(\mathcal{E})$ for delta-coupled detectors given in Eq.~\eqref{eq:classical capacity}, which is  \textit{identical} in form as the channel capacity obtained for \textit{gapless detectors} in \cite{Landulfo2016magnus1}. That is, the functional dependence of $C(\mathcal{E})$ on $\nu_\textsc{b}$ and field commutator $\Delta(f_\textsc{a},f_\textsc{b})$ is completely identical. Their difference lies only ``in the details'': in the delta coupling case, the spacetime smearing $f_j$ is effectively only smeared in the spatial direction, while in gapless case the spacetime smearing includes smooth smearing in the timelike direction. This suggests that the relativistic communication channels provided by both delta coupling model and gapless model are essentially equivalent in their ability to transmit (classical) information. If we have entanglement as a resource, then these two channels are also equivalent in their ability to transmit quantum information since their entanglement-assisted quantum channel capacity $\mathcal{Q}_{ea}(\mathcal{E})\geq \frac{1}{2}C(\mathcal{E})$ \cite{Wilde2013textbook}.

Finally, it is worth stressing that the calculations done so far, including the channel capacity $C(\mathcal{E})$, is very general due to the use of algebraic framework: that is, the results hold in \textit{arbitrary} (globally hyperbolic) curved spacetimes and for any CCR representation associated to quasifree algebraic state $\omega$. In fact, we could modify the channel \eqref{eq: final-state-Bob} straightforwardly to incorporate more general Gaussian states where the one-point function of the field does not vanish.

Below we use the example in flat spacetime to show that just like the gapless detector model, the delta-coupled detectors can also achieve channel capacity arbitrarily close to 1.

\subsection{Example: Minkowski space}\label{subsec: example}

As an example, let us consider relativistic communication mediated by massless scalar field in $(3+1)$-dimensional Minkowski spacetime. The only input we need for the channel capacity is $\nu_\textsc{b}$ and $\Delta(f\ts{a},f\ts{b})$. {For inertial detectors at rest in the Minkowski frame $(t,\bx)$, we do not need Fermi normal coordinates and simply use global Minkowski coordinates.} The spacetime smearing then takes the form $f_j(\tau,\bx) = \lambda_j\eta\delta(\tau-\tau_{j,0})F_j(\bx)$, where $\tau$ is the detector's proper time with $\dd\tau/\dd t = 1$.

We choose a Gaussian function as a spatial profile, i.e.,
\begin{align}
    F_j(\bx)
    &= \dfrac{1}{ ( \sqrt{ \pi } \sigma )^3 } e^{ -|\bx-\bx_{j}|^2/\sigma^2}\,,
\end{align}
where $\sigma$ is a typical width of the Gaussian (both detectors have the same effective size $\sigma$) and $\bx_{j}$ denotes the centre of mass coordinate of each detector. The choice of Gaussian function for the spatial profile for convenience, since one can obtain useful closed-form expressions\footnote{Note that Gaussian profile is not compactly supported and hence strictly speaking our spacetime smearing $f_j\not\in\CS$. That said, since the Gaussian tails are very strongly suppressed it can be regarded as effectively compactly supported (say, within $3.5\sigma$ around its centre). One can always use strictly compactly supported smearing $f_j$ at the expense of slightly more involved computation.}. 

Using Eq.~\eqref{eq: algebraic-norm} and the fact that in Minkowski space we have
\begin{align}
    u_\bk(t,\bx) 
    = \frac{1}{\sqrt{2(2\pi)^3|\bk|}}
    e^{-\ii|\bk|t+\ii\bk\cdot\bx}\,,
\end{align}
so that the (unsmeared) Wightman distribution reads
\begin{align}
    \mathsf{W}(\sx,\sx') = \int\frac{\dd^3\bk}{2(2\pi)^3|\bk|} e^{-\ii|\bk|(t-t')+\ii\bk\cdot(\bx-\bx')}\,,
\end{align}
as a (bi-)distribution. It follows that
\begin{align}
    &||Ef_\textsc{b}||^2 = \mathsf{W}(f_\textsc{b},f_\textsc{b}) \notag\\
    &= \int\dd\tau\,\dd\tau'\dd^3\bx\,\dd^3\bx' \int\frac{\dd^3\bk}{2(2\pi)^3|\bk|} e^{-\ii|\bk|(\tau-\tau')+\ii\bk\cdot(\bx-\bx')} \notag\\
    &\hspace{0.4cm}\times \lambda_\textsc{b}^2\eta^2\delta(\tau-\tau_\textsc{b,0})\delta(\tau'-\tau_\textsc{b,0})F_\textsc{b}(\bx)F_\textsc{b}(\bx')\notag\\
    &= \frac{\tilde\lambda^2_\textsc{b}}{4\pi^2\sigma^2}\,.
\end{align}
It follows from Eq.~\eqref{eq: nub-main} that
\begin{align}
    \nu_\textsc{b} = e^{-2||Ef_\textsc{b}||^2} = \exp
        \kako{
            -\dfrac{ \tilde \lambda_\textsc{b}^2 }{ 2\pi^2 \sigma^2 }
        }\,.
\end{align}
The (unsmeared) causal propagator in Minkowski space $E(\sx,\sx')\equiv E(t,\bx;t',\bx')$ can be computed in two ways: either using the fact that formally it is given in terms of the Wightman distribution as 
\begin{align}
    E(\sx,\sx') = -\ii(\mathsf{W}(\sx,\sx') - \mathsf{W}(\sx',\sx))\,,
    \label{eq: causal-propagator-mink}
\end{align}
or alternatively by direct computation using the fact that formally $E(\sx,\sx')$ is the so-called Pauli-Jordan distribution, i.e., $\ii E(\sx,\sx') = \braket{[\hat\phi(\sx),\hat \phi(\sx)]}$ (see, e.g., \cite{birrell1984quantum} and more explicit computation in the Appendix~C of \cite{Tjoa2021notharvesting}): 
\begin{align}
    E(\sx,\sx') &= \frac{1}{4\pi|\bx-\bx'|}\Bigr[\delta(t-t'+|\bx-\bx'|)\notag\\
    &\hspace{1.8cm} -\delta(t-t'-|\bx-\bx'|)\Bigr]\,.
\end{align}
We will use Eq.~\eqref{eq: causal-propagator-mink} since it will be a straightforward use of Fourier transform. We have
\begin{align}
    &\ii (Ef_\textsc{b})(\sx) \notag \\
    &= \tilde{\lambda}_\textsc{b}\int\dd\tau
    \,\dd^3 \bx'\int\frac{\dd^3\bk}{2(2\pi)^3|\bk|}\delta(\tau'-\tau_{\textsc{b},0})F_\textsc{b}(\bx')\notag\\
    &\hspace{0.4cm} \times \Bigr(e^{-\ii|\bk|(\tau-\tau')+\ii\bk\cdot(\bx-\bx')} - e^{\ii|\bk|(\tau-\tau')-\ii\bk\cdot(\bx-\bx')}\Bigr)\notag\\
    &= \tilde{\lambda}_\textsc{b}\int\frac{\dd^3\bk}{2(2\pi)^3|\bk|}\Bigr[\tilde{F}^*_\textsc{b}(\bk) e^{-\ii|\bk|(\tau-\tau_{\textsc{b},0})}e^{\ii\bk\cdot\bx} - \text{c.c.}\Bigr]\,,
\end{align}
where the Fourier transform reads
\begin{align}
    \tilde{F}_j(\bk) = \int \dd^3\bx\,F_j(\bx)e^{\ii\bk\cdot\bx} \,.
\end{align}
Now the smeared causal propagator reads 
\begin{align}
    &\Delta(f_\textsc{a},f_\textsc{b}) 
    = {-\ii}\int\dd^4\sx \,f_\textsc{a}(\sx)(Ef_\textsc{b})(\sx)\notag \\
    &= {-\ii}
    \tilde{\lambda}_\textsc{a}\tilde{\lambda}_\textsc{b}
    \int\frac{\dd^3\bk}{2(2\pi)^3|\bk|}\Bigr[\tilde{F}_\textsc{a}(\bk)\tilde{F}^*_\textsc{b}(\bk)e^{\ii|\bk|\Delta\tau_\textsc{ab}} - \text{c.c.} \Bigr]\,.
\end{align}
Using the properties of Gaussian function, this integral can be solved exactly to yield 
\begin{align}
    \Delta(f\ts{a},f\ts{b})
    &=  \dfrac{ \tilde \lambda\ts{a} \tilde \lambda\ts{b} }{ 4 \pi^2 L \sigma } \sqrt{ \dfrac{\pi}{2} }
        \kagikako{
            e^{ - \tfrac{ (\Delta \tau_\textsc{ab} - L)^2 }{2\sigma^2} }
            -
            e^{ - \tfrac{ (\Delta \tau_\textsc{ab} + L)^2 }{2\sigma^2} }
        },
\end{align}
where $\Delta \tau_\textsc{ab} \coloneqq \tau\ts{b,0}-\tau\ts{a,0}$ and $L$ is the spatial separation between Alice and Bob. 

\begin{figure}[tp]
    \includegraphics[scale=0.95]{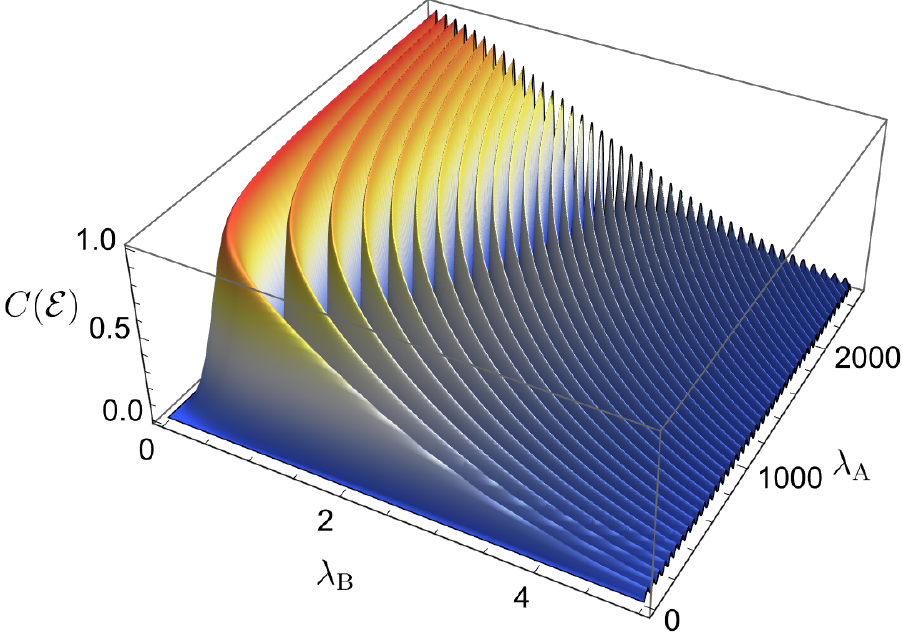}
    \caption{
    The classical channel capacity $C(\mathcal{E})$ as a function of coupling strengths of Alice and Bob's detectors to the field $\lambda_\textsc{a},\lambda_\textsc{b}$. 
    Here the parameters chosen are $\eta/\sigma=1, L/\sigma=6$, and $\Delta\tau_\textsc{ab}/\sigma=6$. We see that $C(\mathcal{E})$ can approach unity if Alice couples much more strongly to the field than Bob.}
    \label{fig: CapacityFig}
\end{figure}

We plot the channel capacity $C(\mathcal{E})$ \eqref{eq:classical capacity} in Fig.~\ref{fig: CapacityFig} as a function of coupling strengths $\lambda\ts{a}$ and $\lambda\ts{b}$. We see that just like the gapless detector case considered in \cite{Landulfo2016magnus1}, delta-coupling can also attain $C(\mathcal{E})\simeq 1$ when Alice couples strongly to the field while Bob interacts weakly. It is important to note that the channel capacity $C(\mathcal{E})$ is maximal when both detectors (assumed to have the same size) are coupled to the field with vastly different coupling strengths, with Alice coupling much more strongly than Bob, i.e., $\lambda_\textsc{a}\gg \lambda_\textsc{b}$ (since the switching strength $\eta_j$ in Eq.~\eqref{eq: switching} is the same). This agrees with the claim in \cite{Simidzija2020transmit} that any decent amount of channel capacity (classical or quantum) for a quantum channel mediated by the field must necessarily be in the nonperturbative regime.

\section{Discussion and outlook}
\label{sec: results}

In this work we revisited the relativistic quantum communication between two UDW detectors interacting locally with a real scalar field via a delta-coupling interaction. Each detector interacts locally with the field very rapidly (effectively instantaneous interaction at a single instant in time), which allows for nonperturbative methods to be employed. We showed that relativistic communication via delta-coupled detectors and gapless detectors are essentially equivalent in the following sense: the quantum channels for both delta-coupling model studied here and gapless detector model studied in \cite{Landulfo2016magnus1} have channel capacity $C(\mathcal{E})$ that obeys the exact same formula,
\begin{align}
    C(\mathcal{E}) &= H\left(\frac{1}{2}+\frac{\nu_\textsc{b}}{2} \bigr|\cos(2\Delta(f_\textsc{a},f_\textsc{b}))\bigr|\right) - H\left(\frac{1}{2}+\frac{\nu_\textsc{b}}{2}\right)\,, \notag
\end{align}
where $\Delta(f_\textsc{a},f_\textsc{b})$ vanishes if the detectors are spacelike separated and $\nu_\textsc{b}$ depends on only the properties of Bob's detectors and the field's state. Furthermore, we generalize the construction to arbitrary initial state of Bob's detector and show that the bound above is optimal and attained only when Bob's detector is initially in a pure state. Consequently, this proves that with shared entanglement as a resource, the \textit{entanglement-assisted} quantum channel capacity, denoted $\mathcal{Q}_{\text{ea}}(\mathcal{E})$, for delta-coupling model can attain its minimum value of $1/2$  (in bits per unit time).

This work constitutes a generalization of delta-coupling UDW model to \textit{arbitrary} (globally hyperbolic) curved spacetimes, as we presented the calculations via algebraic approach in QFT instead of the more conventional calculation in flat spacetime \cite{Gallock2021nonperturbative,Henderson2020Bandlimit,Simidzija2018nogo,Simidzija2020transmit,Koji2020superadditive,sahu2021sabotaging,Henderson2020Bandlimit}). Furthermore, as we do not need to pick a specific CCR representation until the very end, this calculation is much simpler to work with when one wishes to consider different field states. This generalization should prove useful for more general settings where delta coupling remains useful both conceptually and practically, e.g., in spacetimes where the causal propagator can be easily calculated. Finally, this work aims partly to somewhat popularize a more algebraic language within (at least) the RQI community, since AQFT framework is conceptually much cleaner and deals with all CCR representations simultaneously until the very last step.

There are some further investigations that may be of interest following this work. First, all nonperturbative methods in relativistic communication protocol have been mostly focused on scalar fields: in \cite{Simidzija2020transmit} it was shown that Bob's spacetime smearing has to be spherically symmetric to catch all of Alice's signals for transmission of quantum information. In practice, polarization of electromagnetic field may be suitable for encoding qubit information, and perhaps nonperturbative method can shed light on the importance of polarization and directionality in relativistic communication protocols.  Second, it would also be interesting to embed all well-known communication protocols such as teleportation and superdense coding with relativity from first principles: for example, we will have statements such as ``no-communication theorem'' into a simple consequence of the underlying QFT. We leave these further investigations for future work.

\section*{Acknowledgment}

This work was supported in part by the Natural Sciences and Engineering Research Council of Canada. The authors (especially E.T.) thank Finnian Gray for useful discussions on aspects of algebraic QFT. E.T. acknowledges generous support of Mike and Ophelia Lazaridis Fellowship. This work is conducted on the traditional territory of the Neutral, Anishnaabeg, and Haudenosaunee Peoples. The University of Waterloo and the Institute for Quantum Computing are situated on the Haldimand Tract, land that was promised to Six Nations, which includes six miles on each side of the Grand River.

\appendix

\section{Canonical quantization}
\label{appendix: canonical-quantization}

In this section we briefly review canonical quantization procedure and how it connects to the algebraic framework. The key takeaway is that in the standard canonical quantization, the choice of quantization frame and hence a particular representation of the CCR algebra is unavoidable as the creation and annihilation operators require us to single out a preferred Fock space (and hence a vacuum state) of the theory from the outset. In contrast, in algebraic framework one does not need to do this until the very last step, since the algebraic framework deals with all representations of the CCR algebra all at once.

\subsection{Textbook, not manifestly covariant version}

First let us write down the expressions for canonical quantization of a real scalar field that is ``not manifestly covariant'' as is presented in standard textbooks in QFT (see, e.g., \cite{birrell1984quantum}). The idea is that if the spacetime has some sort of time-translation symmetry\footnote{It does not have to be true time-translation symmetry: for instance, in Friedmann-Robertson-Walker (FRW) model of expanding universe, canonical quantization only requires \textit{conformal} timelike Killing vector, which defines the so-called conformal vacuum for the theory. One can also define adiabatic vacuum by relaxing the time translation requirement into the demand that the geometry is ``slowly-varying'' \cite{birrell1984quantum}.}, such as having a timelike Killing vector $\xi$, we can perform the Fourier mode decomposition of the field
\begin{align}
    \phi(\sx) = \int\dd^3\bk\,a_\bk u_\bk(\sx) + a_\bk^* u^*_\bk(\sx)
    \label{eq: fourier-decomposition}
\end{align}
where the modes $\{u_\bk\}$ are eigenmodes of the Klein-Gordon operator $P$ defined by Eq.~\eqref{eq: KGE} and they are \textit{positive-frequency} with respect to $\xi^a$: that is, $u_\bk$ satisfies the eigenvalue equation
\begin{align}
    \ii\mathcal{L}_\xi u_\bk(\sx) = \omega_\bk u_\bk(\sx)\,,\quad \omega_\bk>0\,.
\end{align}
where $\mathcal{L}_\xi$ is the Lie derivative with respect to $\xi$. With the mode decomposition \eqref{eq: fourier-decomposition}, we promote the field into an operator-valued distribution
\begin{align}
    \hat\phi(\sx) = \int\dd^3\bk\,\hat a_\bk u_\bk(\sx) + \hat a_\bk^\dagger u^*_\bk(\sx)\,,
\end{align}
where the operators $\hat a_\bk,\hat a_\bk^\dagger$ are now ladder operators satisfying the canonical commutation relations (CCR) $[\hat a_\bk,\hat a_{\bk'}^\dagger] = \delta^3(\bk-\bk') \openone$. The vacuum state $\ket{0}$ is then an element of the Fock space such that $\hat a_\bk\ket{0}=0$ for all $\bk$. One can check that if this quantization is performed in the quantization frame $(t,\bx)$ based on some constant-$t$ foliation $\R\times\Sigma_t$ where $\Sigma_t$ is a spacelike Cauchy surface, then this reproduces the usual equal-time CCR
\begin{align}
    [\hat\phi(t,\bx),\hat\pi(t,\bx')] &= \ii\delta^3(\bx-\bx') \openone \,,\\
    [\hat\phi(t,\bx),\hat\phi(t,\bx')] &= [\hat\pi(t,\bx),\hat\pi(t,\bx')] = 0\,,
\end{align}
where the canonical conjugate momentum is defined (in curved spacetime) by
\begin{align}
    \pi(t,\bx) = \sqrt{h}n^a\nabla_a\phi(t,\bx)
\end{align}
where $h=\det h_{ij}|_{\Sigma_t}$ is the determinant of the induced metric $h_{ij}(\bx)$ on $\Sigma_t$. In Minkowski space and taking $\Sigma_t$ to be the usual constant-$t$ surfaces, this reduces to $\pi = \partial_t\phi$.

The problem with canonical quantization is that there are many unitarily inequivalent representations of the CCR algebra \cite{wald1994quantum,birrell1984quantum}. For example, in Minkowski space the quantization based on the modes $\{ u_\bk(t,\bx) \}$  which are positive frequency with respect to time-translation Killing vector $\xi = \partial_t$ defines the so-called \textit{Minkowski vacuum} $\ket{0_\textsc{M}}$. However, we could try to quantize the field in \textit{Rindler frame} $(\eta,\tilde{\bx})$ associated to constantly accelerating observers (say, in $x$-direction) and obtain another mode decomposition
\begin{align}
     \hat\phi(\eta,\tilde{\bx}) = \int\dd^3\bk\,\hat b_\bk v_\bk(\eta,\tilde\bx) + \hat b_\bk^\dagger v^*_\bk(\eta,\tilde\bx)\,.
\end{align}
In this case, the modes $\{v_\bk(\eta,\tilde\bx)\}$ are eigenmodes of Klein-Gordon operator $P$ that is positive frequency with respect to the \textit{boost} Killing vector $\mathsf{K}=x\partial_t +t\partial_x$:
\begin{align}
    \ii\mathcal{L}_\mathsf{K} v_\bk = \omega_\bk v_\bk\,,\quad \omega_\bk>0\,.
\end{align}
The ladder operators $\hat b_\bk,\hat b_\bk^\dagger$ also obeys the usual CCR \mbox{$[\hat b_\bk,\hat b_{\bk'}^\dagger] = \delta^3(\bk-\bk')\openone$}. The \textit{Rindler vacuum} $\ket{0_\textsc{R}}$ is defined by $\hat b_\bk\ket{0_\textsc{R}}=0$ for all $\bk$. The two vacua are not unitarily equivalent because the number operators of one quantization does not register zero for another:
\begin{align}
    \braket{0_\textsc{R}|\hat a_\bk^\dagger \hat a_\bk^{\phantom{\dagger}} |0_\textsc{R}}\neq 0\,,\quad \braket{0_\textsc{M}|\hat b_\bk^\dagger \hat b_\bk^{\phantom{\dagger}} |0_\textsc{M}} \neq 0\,,
\end{align}
hence the two observers disagree on the particle content of each other's vacua.

\subsection{Manifestly covariant version}

The equal-time CCR is not manifestly covariant as it singles out a preferred time direction. The way to do this more covariantly is by first consider the full complexified space of solutions to Klein-Gordon equation, denoted $\Sol_\C(\M)$, and make it into a symplectic vector space by giving it a symplectic two-form 
\begin{align}
    \sigma(\varphi_1,\varphi_2) \coloneqq \int_{\Sigma_t}\dd\Sigma n^a[\varphi_2\nabla_a\varphi_1-\varphi_1\nabla_a\varphi_2]\,,
\end{align}
where $\dd\Sigma = \dd^3\bx\sqrt{h}$ is the proper volume element of the Cauchy surface. We can use this to define the \textit{Klein-Gordon inner product}
$(\cdot,\cdot)_\textsc{kg}: \Sol_\C(\M)\times \Sol_\C(\M) \to \C$ 
\begin{align}
    (\varphi_1,\varphi_2)_\textsc{kg} = -\ii\sigma(\varphi^*_1,\varphi_2)\,.
\end{align}
For the eigenmodes $\{u_\bk\}$, this gives the normalization and orthogonality condition in Eq.~\eqref{eq: KG-normalization}. 

In the previous subsection, the Fock space is essentially taken by granted as an ``analogy'' with (infinitely many) harmonic oscillators with frequency $\omega_\bk$. However, in the covariant version this construction can be made more explicit. The idea is that $\Sol_\C(\mathcal{M})$ equipped with Klein-Gordon inner product is not a Hilbert space because it is not positive definite (essentially due to negative-frequency modes). Therefore, we should be able to single out just the positive-frequency part spanned by $\{u_\bk\}$ only (without $\{u_\bk^*\}$) and make that into a Hilbert space. Indeed, it can be shown that this can be done \cite{wald1994quantum}: there exists $\mathcal{H}\subset \Sol_\C(\M)$ such that
\begin{enumerate}[label=(\alph*),leftmargin=*]
    \item $(\mathcal{H},(\cdot,\cdot)_{\textsc{kg}})$ {is} a Hilbert space;
    
    \item $\Sol_\C(\M)\cong \mathcal{H}\oplus\overline{\mathcal{H}}$, where $\overline{\mathcal{H}}$ is the \textit{complex conjugate} Hilbert space (not dual space $\mathcal{H}^*$!) of $\mathcal{H}$.
\end{enumerate}
It follows that if $u\in \overline{\mathcal{H}}$ and $v\in {\mathcal{H}}$ and $(u,v)_\textsc{kg}=0$. The Fock space of the real scalar field is then given by the symmetrized direct sum
\begin{align}
    \mathfrak{F}_s(\mathcal{H})\coloneqq \bigoplus_{n=0}^\infty \mathcal{H}^{\otimes_s n} = \C\oplus \mathcal{H} \oplus (\mathcal{H}\otimes_s\mathcal{H}) \oplus ...
\end{align}
where the subscript in $\otimes_s$ denotes the symmetrization since the field is bosonic. The Hilbert space $\mathcal{H}$ is called the \textit{one-particle Hilbert space} of the field. The smeared field operator $\hat\phi(f)$ then acts on the Fock space $\mathfrak{F}_s(\mathcal{H})$. 

In this language, the (unsmeared) field operator is usually written as \cite{wald1994quantum}
\begin{align}
    \hat \phi(\sx) &= \int \dd^3\bk\,\left[ \hat a(u_\bk^*) u_\bk(\sx) + \hat a^\dagger(u_\bk) u^*_\bk(\sx)\right]\,,
\end{align}
where the annihilation and creation operators are really viewed as \textit{linear operators} taking elements of $\overline{\mathcal{H}}$  and $\mathcal{H}$ respectively to a linear operator acting on $\mathfrak{F}_s(\mathcal{H})$. Formally, we write this as
\begin{align}
    \hat a&: \overline{\mathcal{H}}\to \text{End}(\mathfrak{F}_s(\mathcal{H}))\,,\quad\hat a^\dagger: {\mathcal{H}}\to \text{End}(\mathfrak{F}_s(\mathcal{H}))\,.
\end{align}
Note that the operators $\hat a(u_\bk^*),\hat a^\dagger(u_\bk): \mathfrak{F}_s(\mathcal{H})\to \mathfrak{F}_s(\mathcal{H})$ are what gives the ``shorthand'' $\hat a_\bk^{\phantom{\dagger}},\hat a_\bk^\dagger$ in the standard version. The covariant CCR then reads
\begin{align}
    [\hat a(u^*),\hat a^\dagger(v)] = (u,v)_\textsc{kg}\openone\,,\quad u,v\in\mathcal{H}\,.
\end{align}
The field operator with these ladder operators then define a representation of the CCR algebra and the vacuum state is the vector $\ket{0}\in\mathfrak{F}_s(\mathcal{H})$ with the property
\begin{align}
    \hat a(u)\ket{0} = 0\quad \forall u\in \overline{\mathcal{H}}\,.
\end{align}
Finally, the smeared field operator can be written using the above as \cite{wald1994quantum}
\begin{align}
    \hat\phi(f) &= \ii\bigr[\hat a((KEf)^*) - \hat a^\dagger (KEf)\bigr] \,,
    \label{eq: smeared-field-KEf}
\end{align}
and we recover the standard unsmeared expression using Eq.~\eqref{eq: KEf}.

\textbf{Remark:} Note that different sources use different notation for the creation and annihilation operators, which  affects the presentation of the smeared field operator in Eq.~\eqref{eq: smeared-field-KEf}. The convention here is basically that of Wald \cite{wald1994quantum} (also used by \cite{Landulfo2016magnus1}), where $\hat a$ is viewed as a \textit{linear} map acting on $\overline{\mathcal{H}}$, so it reads $a(u_\bk^*)$. In contrast, in \cite{fewster2020algebraic} they write $\hat a(u_\bk)$ so they view $\hat a$ as \textit{antilinear} map on $\mathcal{H}$. In \cite{Khavkhine2015AQFT}, they take $\hat a$ as a map acting on $\Sol_\R(\M)$ instead of $\mathcal{H}$ or $\overline{\mathcal{H}}$, so they write $\hat a(Ef)$ for the annihilation operators (since any solution can be written as $Ef$ for some $f\in\CS$). Therefore, we have for instance (in \cite{Khavkhine2015AQFT} and \cite{fewster2020algebraic} respectively)
\begin{align}
    \hat\phi(f) &= \hat a(Ef) + \hat a^\dagger (Ef)\,,\\
    \hat\phi(f) &= a(KEf)+a^\dagger(KEf)\,,
\end{align}
In these versions, the complex number $\ii$ in Eq.~\eqref{eq: smeared-field-KEf} is effectively absorbed into the definition of $\hat a,\hat a^\dagger$.

\begin{center}
    ------------------------------------------------------------
\end{center}

It is now clear that the reason why the algebraic approach is preferable in some sense is because one does not need to pick a preferred representation of the CCR algebra, hence the vacuum state, until the very end; the canonical quantization requires this choice very early because one needs the creation and annihilation operators. One can see that in our calculation of channel capacity \eqref{eq:classical capacity}, we do not even refer to any representation until when we need to compute explicitly the quantities such as $\nu_\textsc{b}$ and $\nu_\textsc{ab}^\pm$ in Eq.~\eqref{eq: nuj}-\eqref{eq: nuAB}. We only used the properties of the Weyl algebra to get to that point. 

At this final step in calculating $\nu_j,\nu_{AB}^\pm$, we do need to specify the algebraic state and representation of the CCR algebra because the norm-squared $||Ef||^2$ depends on the choice of $\mu$ (i.e., the vacuum state). This manifests in the need to compute $\mathsf{W}(f,f)$, which depends on $\{u_\bk\}$ (which implicitly defines the ladder operators). Changing the representation only affects this part of the calculation, thus making the algebraic approach much more efficient and cleaner, instead of redoing canonical quantization from scratch.

\section{Calculations of the constants $\gamma_{ijkl}$}
\label{appendix: gammas}

In this section we show the calculation for the constants $\gamma_{ijkl}$ that we need in this work. First, given some smooth compactly supported function $f\in \CS$, the smeared field operator 
\begin{align}
    \hat Y_j \equiv \hat \phi(f_j) = \int \dd V\,f_j(\sx)\hat \phi(\sx)\,,
\end{align}
can be exponentiated to give a bounded operator $e^{\pm \ii\hat Y}$ that is an element of the Weyl algebra $\W(\M)$. In the case of the delta coupling model, we can choose this to be the factorized smearing $f_j(\sx) = \lambda_j\chi_j(\tau) F_j(\bx)$, but in what follows we do not need to make this choice.
Thus we can write
\begin{align}
    W(\pm Ef_j) = e^{\pm \ii \hat Y_j}\,.
\end{align}
It follows that 
\begin{align}
    \cos \hat Y_j &= \frac{e^{\ii\hat Y}+e^{-\ii\hat Y}}{2} = \frac{W(Ef_j)+W(-Ef_j)}{2}\,,\\
    \sin \hat Y_j &= \frac{e^{\ii\hat Y}-e^{-\ii\hat Y}}{2\ii} = \frac{W(Ef_j)-W(-Ef_j)}{2\ii}\,.
\end{align}
Now, using the Weyl relations \eqref{eq: Weyl-relations} and the shorthand $\Delta_{\textsc{ab}}\equiv \Delta(f_\textsc{a},f_\textsc{b})$, it can be shown that
\begin{subequations}
\begin{align}
    \gamma_{cccc} &= \frac{1+\nu_\textsc{a}+\nu_\textsc{b}\cos(2\Delta_{\textsc{ab}})}{4} + \frac{\nu_\textsc{ab}^++\nu_\textsc{ab}^-}{8}\,,\\
    \gamma_{ssss} &= \frac{1-\nu_\textsc{a}-\nu_\textsc{b}\cos(2\Delta_{\textsc{ab}})}{4} + \frac{\nu_\textsc{ab}^++\nu_\textsc{ab}^-}{8}\,,\\
    \gamma_{cssc} &= \frac{1-\nu_\textsc{a}+\nu_\textsc{b}\cos(2\Delta_{\textsc{ab}})}{4} - \frac{\nu_\textsc{ab}^++\nu_\textsc{ab}^-}{8}\,,\\
    \gamma_{sccs} &= \frac{1+\nu_\textsc{a}-\nu_\textsc{b}\cos(2\Delta_{\textsc{ab}})}{4} - \frac{\nu_\textsc{ab}^++\nu_\textsc{ab}^-}{8}\,,\\
    \gamma_{scsc} &=  \frac{-\ii\nu_\textsc{b}\sin(2\Delta_{\textsc{ab}})}{4} - \frac{\nu_\textsc{ab}^+-\nu_\textsc{ab}^-}{8}\,,
    \label{eq: b5e}\\
    \gamma_{sscc} &= \frac{\ii\nu_\textsc{b}\sin(2\Delta_{\textsc{ab}})}{4} - \frac{\nu_\textsc{ab}^+-\nu_\textsc{ab}^-}{8}\,,
\end{align}
\end{subequations}
where
\begin{align}
    \nu_{j} &= \omega\bigr(W(E(2f_j))\bigr) = e^{-2||Ef_j||^2}\,,\label{eq: nuj}\\
    \nu_{\textsc{ab}}^\pm &= \omega\bigr(W(E(f_\textsc{a}+f_\textsc{b}))\bigr) = e^{-2||E(f_\textsc{a}\pm f_\textsc{b})||^2}\,,
    \label{eq: nuAB}
\end{align}
with the norm computed using Eq.~\eqref{eq: norm-Ef} which we write here again for convenience:
\begin{align}
    ||Ef||^2 &= \mathsf{W}(f,f) = \braket{KEf,KEf}_\mathcal{H} \\
    &= \int \dd V \dd V' u_\bk (\sx)u_\bk^*(\sx')f(\sx)f(\sx')\,,
\end{align}
where $\dd V=\dd^4\sx\sqrt{-g}$ is the invariant volume element. 

Let us demonstrate this for $\gamma_{scsc}$ and the rest proceeds analogously. We first write
\begin{align}
    \gamma_{scsc} &= \omega(\sin\hat Y_\textsc{a}\cos\hat Y_\textsc{b}\sin\hat Y_\textsc{b}\cos\hat Y_\textsc{a}) \,. 
\end{align}
We compute $\sin\hat Y_\textsc{a}\cos\hat Y_\textsc{b}$ using Weyl relations. Using the shorthand $W_j=W(Ef_j)$ ($j=\textsc{a,b}$) and $W_{\pm A\pm B}\equiv W(\pm Ef_\textsc{a} \pm Ef_\textsc{b})$,  we have
\begin{align}
    &-4\ii\sin\hat Y_\textsc{a}\cos\hat Y_\textsc{b} \notag\\
    &= (W_{A} - W_{-A})(W_{B}+W_{-B}) \notag\\
    &= W_A W_B + W_A W_{-B} - W_{-A} W_{B} - W_{-A} W_{-B} \notag\\
    &= W_{A+B} e^{\ii\Dab/2} + W_{A-B}e^{-\ii\Dab/2} \notag\\
    &\hspace{0.4cm}  - W_{-A+B}e^{-\ii\Dab/2} -  W_{-A-B} e^{\ii\Dab/2}\,, 
\end{align}
where the Weyl relations have been used in the third equality. 
Therefore, we get
\begin{align}
    &-16\gamma_{scsc} = -16\sin\hat Y_\textsc{a}\cos\hat Y_\textsc{b} \sin\hat Y_\textsc{b}\cos\hat Y_\textsc{a} \notag\\
    &= \bigr(W_{A+B} e^{\ii\Dab/2} + W_{A-B}e^{-\ii\Dab/2} \notag\\
    &\hspace{0.4cm}  - W_{-A+B}e^{-\ii\Dab/2} -  W_{-A-B} e^{\ii\Dab/2}\bigr)\notag\\
    &\hspace{0.4cm} \times \bigr( W_{B+A} e^{-\ii\Dab/2} + W_{B-A}e^{\ii\Dab/2} \notag\\
    &\hspace{0.4cm}  - W_{-B+A}e^{\ii\Dab/2} -  W_{-B-A} e^{-\ii\Dab/2}\bigr)\,,
\end{align} 
where we used the fact that $\Delta(f_\textsc{a},f_\textsc{b}) = -\Delta(f_\textsc{b},f_\textsc{a})$. Expanding and using the Weyl relations, we get
\begin{align}
    -16\gamma_{scsc} 
    &= W_{2A+2B} + W_{2B}e^{2\ii\Dab} - W_{2A} - W_{0} \notag \\
    & + W_{2A} + W_0 - W_{2A-2B} -W_{-2B} e^{-2\ii\Dab} \notag\\
    & - W_{2B}e^{-2\ii\Dab} - W_{2B-2A} + W_0 + W_{-2A} \notag\\
    & - W_0 -W_{-2A} + W_{-2B}e^{2\ii\Dab} + W_{-2A-2B}\,.
\end{align}
Here, we can use the fact that $W(2Ef) = ||Ef||^2 = W(-2Ef)$ and that $W_0 \equiv W(0) = 1$. Hence, using the Weyl relations and Eqs.~\eqref{eq: nuj}-\eqref{eq: nuAB}, we get
\begin{align}
    -16\gamma_{scsc} &= 2\nu_\textsc{ab}^+ -2\nu_\textsc{ab}^- + 4\ii \nu_{\textsc{b}}\sin(2\Dab)\,.
\end{align}
Dividing both sides by $-16$ gives precisely Eq.~\eqref{eq: b5e}. The rest of the $\gamma_{ijkl}$ proceeds in similar fashion.

It is worth noting that this calculation is valid in \textit{arbitrary} curved spacetime so long as the UDW detector model makes sense. This is one of the big advantages we gain by working with algebraic QFT language.

\bibliography{rqc}

\end{document}